\documentclass[11pt]{article} 
\usepackage{amsfonts}
\usepackage[T1]{fontenc}
\usepackage{bm}
\usepackage{amsmath}  
\usepackage{graphicx} 
\usepackage{color}

\newcommand{\be}{ \begin{equation}}
\newcommand{\ee}{\end{equation}} 
\begin{document} 
\def\theequation{\arabic{section}.\arabic{equation}} 
\begin{titlepage} 
\title{When Painlev\'e-Gullstrand coordinates fail} 

\author{Valerio Faraoni and Genevi\`eve Vachon\\ \\ 
{\small Department of Physics \& Astronomy, Bishop's University}\\
{\small 2600 College Street, Sherbrooke, Qu\/'ebec, Canada J1M~1Z7}
 } 

\date{} 
\maketitle 
\thispagestyle{empty} 
\vspace*{1truecm} 
\begin{abstract} 

Painlev\'e-Gullstrand coordinates, a very useful tool in spherical horizon 
thermodynamics, fail in anti-de Sitter space and in the inner region of 
Reissner-Nordstr\"om. We predict this breakdown to occur in any region 
containing negative Misner-Sharp-Hernandez quasilocal mass because of 
repulsive gravity stopping the motion of PG observers in radial free fall 
with zero initial velocity. PG coordinates break down also in the static 
Einstein universe for completely different reasons. The more general 
Martel-Poisson family of charts, which normally has PG coordinates as a 
limit, is reported for static cosmologies (de Sitter, anti-de Sitter and 
the static Einstein universe).

\end{abstract} 
\vspace*{1truecm} 

%
\end{titlepage}


\def\theequation{\arabic{section}.\arabic{equation}}

\section{Introduction}
\label{sec:1}
\setcounter{equation}{0}

Black hole thermodynamics is an important area of modern theoretical 
physics linking quantum processes and classical gravity. The 
thermodynamics of stationary horizons is well developed but, when horizons 
become dynamical ({\em i.e.}, timelike or spacelike apparent/trapping 
horizons instead of null event horizons \cite{mybook}), our undertanding 
their thermodynamics drops dramatically. A valuable tool to obtain the 
Hawking temperature of time-dependent horizons is the tunneling formalism 
pioneered by Parikh and Wilcek \cite{ParikhWilczek}, which uses 
Painlev\'e-Gullstrand (PG) coordinates \cite{Painleve,Gullstrand} 
penetrating the horizon (see \cite{Vanzo} for a review). A chacteristic 
feature of PG coordinates is that 
the 3-dimensional spatial sections of spacetimes foliated by these 
coordinates are flat. PG coordinates constitute a very useful chart also 
in other problems in classical and quantum gravity where 
Schwarzschild-like (or ``curvature'') coordinates fail 
\cite{KrausWilczek,Medved, 
HawkingHunter,Kanai,AdlerBjorkenLiu,Finch,NielsenVisser, 
AbreuVisser,GoodingUnruh, MacLaurin,Boonsermetal,Murchada, Viqar, 
HusainWinkler,  
BeigSiddiqui,Akbar, Lisle,mybook, Kobayashi}.  
Therefore, from the point of view of tool building and in view of their 
many applications, it is important to have a complete understanding of PG 
coordinates.

It is sometimes stated explicitly in the literature that all static and  
spherically symmetric spacetimes admit PG coordinates,  but there are such 
situations 
of physical interest where PG coordinates fail,  
and one must resort to less optimal tools. These situations include the 
Schwarzschild-anti de Sitter family of black holes \cite{}. It turns out 
that the problem is not the black hole itself, but rather the anti-de 
Sitter background in which the latter is embedded. It has been pointed out 
that PG coordinates cannot be constructed for anti-de Sitter space 
\cite{LinSoo}. On the other hand, a recipe that is quite general to 
construct PG coordinates in spherically symmetric spacetimes exists 
\cite{MartelPoisson, NielsenVisser}. Here we explain the difficulties with 
anti-de Sitter space and with many other spherical geometries (static or 
not), from both the mathematical and the physical points of view.

PG coordinates \cite{Painleve,Gullstrand} are just a special case 
(corresponding to a special value of the parameter) of the more general 
one-parameter Martel-Poisson family of charts. Therefore, we first discuss 
the general Martel-Poisson family. Not surprisingly, the prototypical 
geometry in which both PG and Martel-Poisson coordinates were originally 
introduced is the Schwarzschild spacetime, to which one often refers to 
gain physical intuition for more general situations, expecially in black 
hole thermodynamics. We will abide to this unwritten rule and use the 
Schwarzschild spacetime to shed light on different geometries.

Motivated by the puzzle with anti-de Sitter space, we consider static 
cosmological metrics, including de Sitter and anti-de Sitter space and the 
Einstein static universe.  The Martel-Poisson coordinates 
\cite{MartelPoisson} for the Schwarschild spacetime are based on radial 
timelike geodesics and they use as time coordinate the proper time of 
observers in radial free fall. Martel and Poisson \cite{MartelPoisson} 
give a detailed mathematical construction and physical interpretation for 
the Schwarzschild geometry, and also outline how to construct similar 
coordinates for generic static spherically symmetric spacetimes 
\cite{MartelPoisson}. In the realm of cosmology, de Sitter and anti-de 
Sitter spaces and the Einstein static universe are spherically symmetric 
and locally static, like Schwarschild. The main difference with respect to 
Schwarzschild in the construction of Martel-Poisson coordinates is that 
one needs to consider {\em outgoing} massive observers in radial motion 
starting from a centre instead of {\em ingoing} observers falling radially 
from infinity. This fact leads to some differences with respect to the 
Martel-Poisson treatment, which are highlighted here.

Let us begin by reviewing Martel-Poisson (and their special PG subcase) 
coordinates for the Schwarzschild spacetime\footnote{We follow the 
notation of Ref.~\cite{Wald}.} \cite{MartelPoisson} 
\be
ds^2 =-\left( 1-\frac{2M}{r} \right)dt^2 + \frac{dr^2}{1-2M/r} +r^2 
d\Omega_{(2)}^2  \equiv =-fdt^2 + \frac{dr^2}{f} +r^2 
d\Omega_{(2)}^2 \,,\label{Schwarzschild}
\ee
where $d\Omega_{(2)}^2=d\vartheta^2 +\sin^2 \vartheta \, d\varphi^2$ is 
the line element on the unit 2-sphere. The Martel-Poisson family of charts 
is 
parametrized by a parameter $p$ spanning the range $ 0<p\leq 1$. 
The family includes the Painlev\'e-Gullstrand coordinates (for $p=1$) and 
the Eddington-Finkelstein (EF) coordinates  (obtained in the limit 
$p\rightarrow0$). Let $\tau$ denote the proper time 
along radial timelike geodesics and $u^a $ be the 
particle four-velocity, which is related to its coordinate three-velocity 
by
\be
u^{\mu}\equiv \frac{dx^{\mu}}{d\tau}= \frac{dx^{\mu}}{dt} \, 
\frac{dt}{d\tau} = \gamma \, \frac{dx^{\mu}}{dt} =
\gamma \left( 1, \frac{d\vec{x}}{dt} \right) \equiv \left( \gamma, \gamma 
\vec{v} \right) \,,
\ee
where $\vec{v}$ is the coordinate 3-velocity and $\gamma( v) =\left( 
1-v^2\right)^{-1/2}$ is the Lorentz factor. 
The  Martel-Poisson coordinates for the Schwarzschild spacetime are 
based on {\em ingoing} freely falling observers with purely radial 
velocity. The energy $E$ of  a particle of mass $m$ is conserved along the 
geodesics, 
\be
m u_c t^c = -E \,,
\ee
where $t^c \equiv \left( \partial /\partial t \right)^c $ is the 
timelike Killing vector. Introducing the particle energy per unit mass 
$\bar{E} \equiv E/m$, it is 
\be
\frac{dt}{d\tau} = \frac{ \bar{E}}{f} \,.\label{conservation}
\ee
The normalization $u_c u^c =-1$ reads 
\be
-f \left( \frac{dt}{d\tau}\right)^2 
+ \frac{1}{f} \left( \frac{dr}{d\tau}\right)^2 = -1 \,;
\ee
Eq.~(\ref{conservation}) then yields
\be
\left( \frac{dr}{d\tau}\right)^2=\bar{E}^2 -f
\ee
and 
\be
\frac{dr}{d\tau} = - \gamma v = - \sqrt{ \bar{E}^2-f}  \label{boh}
\ee
with $v\equiv \left| \vec{v} \right|$.
At $r=\infty $, the coordinate time $ t$  of static observers 
coincides with the proper time $\tau$ along the geodesics and 
Eq.~(\ref{boh}) reads
\be
v_{\infty} \equiv \left|\frac{dr}{dt} \right|_{\infty} =  \sqrt{ 
\bar{E}^2-1} \,.
\ee
The parameter $p$ is defined as   
\be
p \equiv \frac{1}{\bar{E}^2} = 1-v_{\infty}^2  \,.
\ee
The Martel-Poisson coordinates are defined by 
\be
d\bar{t} = dt + \frac{\sqrt{ 1-pf}}{f}\,  dr 
\ee
or, in integral form,
\be
\bar{t} = t + \int \frac{\sqrt{ 1-pf}}{f}\,  dr 
\ee
and the Schwarzschild line element becomes
\be
ds^2 = -fd\bar{t}^2 \pm 2 \sqrt{1-pf} \, d\bar{t}dr +pdr^2 +r^2 
d\Omega_{(2)}^2 \,.
\ee
For $p=1$, which corresponds to observers infalling radially from 
infinity with zero initial velocity $v_{\infty}=0$, the Martel-Poisson 
coordinates reduce to the more familiar PG 
coordinates in which the Schwarzschild line element assumes the form
\be
ds^2 = - \left(1-\frac{2M}{r} \right)d\bar{t}^2 + 2 \sqrt{\frac{2M}{r}} 
d\bar{t}dr +dr^2 +r^2 \Omega_{(2)}^2 
\ee
and 
\be 
\bar{t} = t+4M\left( \sqrt{\frac{r}{2M} } +\ln \sqrt{ \left| 
\frac{\sqrt{r/(2M)} -1}{\sqrt{r/(2M)} +1} \right| } \right)\,.
\ee
In the limit $p\rightarrow 0$, EF coordinates 
\cite{Eddington,Finkelstein} are 
obtained \cite{MartelPoisson}. First, one introduces the tortoise 
coordinate
\be
r^* \equiv r+2M \ln \left| \frac{r}{2M} -1\right| =r+2M \ln \left| 
\frac{r}{2M} 
\left(1-\frac{2M}{r} \right)\right| \,,
\ee
whose differential satisfies
\be
dr^* = \frac{dr}{1-2M/r} \,.
\ee
The null coordinates $\left( u,v\right)$ are introduced by
\begin{eqnarray}
du & \equiv & dt-dr^* \,,\\
&&\nonumber\\
dv & \equiv & dt+dr^* \,,
\end{eqnarray}
and $dt=du+dr^*=dv-dr^*$. 
Ingoing ($-$) EF coordinates \cite{Eddington,Finkelstein} use the advanced 
time $v$ and the 
Schwarzschild line element~(\ref{Schwarzschild}) is written as
\be
ds^2_{(-)} = -\left(1-\frac{2M}{r}\right)  dv^2 +2dvdr +r^2 
d\Omega_{(2)}^2 \,,
\ee
while outgoing ($+$) coordinates use the retarded time $u$, with
\be
ds^2_{(+)} = -\left(1-\frac{2M}{r}\right)  du^2 -2dudr +r^2 
d\Omega_{(2)}^2 \,.
\ee

Martel and Poisson \cite{MartelPoisson} proceed to generalize the 
construction of their coordinates to more general static and spherically 
symmetric spacetimes
\be 
ds^2=-\mbox{e}^{-2\Phi} fdt^2+\frac{dr^2}{f} + r^2 d\Omega_{(2)}^2
\ee
with $\Phi-\Phi(r), f=f(r)$. By redefining the time coordinate according 
to 
\be
T=t+\int \frac{ \sqrt{\mbox{e}^{2\Phi} -p f}}{f} \, dr \,,
\ee
the line element becomes
\be
ds^2 =-\mbox{e}^{-2\Phi} f dT^2 +2\mbox{e}^{-2\Phi}\sqrt{ \mbox{e}^{2\Phi} 
- pf}\, dTdr +p\mbox{e}^{-2\Phi}dr^2 +r^2 d\Omega_{(2)}^2   
\ee
PG coordinates are obtained in the imit $p\rightarrow 1$ and 3-surfaces of 
constant $T$ are flat \cite{MartelPoisson}. 

Although not noted in \cite{MartelPoisson}, this construction breaks down 
when the argument of the square root becomes negative. This fact was noted 
in \cite{LinSoo} for anti-de Sitter space and for the inner region of the 
Reissner-Nordstr\"om metric.  These two regions have in common a negative 
quasilocal mass. We  show in Sec.~\ref{sec:4} that, whenever this 
happens, PG coordinates cannot be introduced and we highlight the   
physical reason: since gravity becomes repulsive, a massive test particle 
with zero initial velocity cannot overcome this repulsion and it 
cannot even begin to travel radially along a radial timelike geodesic. For 
illustration, we refer to the case of the Schwarschild naked singularity 
(which is illuminating as usual) and then revert to anti-de Sitter space.

\section{Martel-Poisson family of charts for de Sitter space}
\setcounter{equation}{0}
\label{sec:2}

Various coordinate charts in de Sitter space are reviewed in 
Refs.~\cite{EriksenGron,Pascu}. Here we limit ourselves to 
the Poisson-Martel family of charts  and its limits.
Begin from the de Sitter line element in Schwarzschild-like (or curvature) 
coordinates
\be
ds^2 =-\left(1-H^2R^2\right)dT^2 +\frac{dR^2}{1-H^2R^2} +R^2 
d\Omega_{(2)}^2 \equiv -fdT^2 +\frac{dR^2}{f} +R^2 
d\Omega_{(2)}^2  \,, \label{dSstatic}
\ee
where $H$ is constant and the line element is locally static in 
the region $0\leq R\leq H^{-1}$,  
and define a new time coordinate $\bar{T}$ by\footnote{Changing the sign 
of the term in $dR$ in Eq.~(\ref{dSdT}) leaves $d\bar{T}$ an exact 
differential and  changes the sign of $dR/d\tau$ and 
switches from ingoing to outgoing geodesics and {\em vice-versa}. Since in 
the following  we are already considering both outgoing and ingoing 
geodesics, we do not need to include a $\pm $ sign here.}
   \be
d\bar{T}= dT + \frac{ \sqrt{1-pf}}{f} \, dR \,, \label{dSdT}
\ee
where $p$ is a parameter labelling different charts.  Clearly, the 
components of the differential $d\bar{T}=c_1 dT +c_2 dR$ satisfy $
\partial c_1 / \partial R = \partial c_2 / \partial T$ and $d\bar{T}$ is 
exact.

To see the physical meaning of $p$, write the equation of {\em 
outgoing} ($\dot{R}>0$) radial timelike geodesics  
\be
\frac{ds^2}{d\tau^2} = -f \left( \frac{dT}{d\tau} \right)^2 +\frac{1}{f} 
\, \left( \frac{dR}{d\tau} \right)^2 =-1 \,,\label{dStimelikeradial1}
\ee
where $\tau$ is the proper time along timelike geodesics. Since the de 
Sitter metric is locally static the energy is conserved along geodesics. 
If 
$T^a= \left( \partial/\partial T \right)^a $ denotes the 
timelike Killing vector and $p^{c}=m u^c$ the four-momentum of  a 
massive particle of mass $m$ and 4-velocity $u^c$, then $
p_a T^a= -E $ 
is constant  along the geodesic. Using the energy per unit mass $\bar{E} 
\equiv E/m$, we have $ 
u^0= dT/d\tau = \bar{E}/f$. Substituting into the radial timelike geodesic 
equation~(\ref{dStimelikeradial1}) yields
\be
\left( \frac{dR}{d\tau} \right)^2= \bar{E}^2-f
\ee
and 
\be
\frac{dR}{d\tau} = \pm \sqrt{ \bar{E}^2-f}
\ee
with the upper sign for outgoing and the lower sign for ingoing 
geodesics. 
Introduce the parameter $p\equiv 1/\bar{E}^2 $ 
and  consider the radial component of the four-velocity of the 
massive particle
\be
\frac{dR}{d\tau}=\frac{dR}{dt}\, \frac{dt}{d\tau} =\pm \gamma(v) \, v 
=\pm \frac{v}{\sqrt{1-v^2}} =\pm \sqrt{\bar{E}^2-f} \,,
\ee
where $\gamma(v)$ is the Lorentz factor and $v=|\vec{v}|$ is the 
magnitude of the coordinate 3-velocity. 

At $R=0$ we have 
\be
\left| \frac{dR}{d\tau}\Big|_{R=0} \right| = \frac{v_0}{\sqrt{1-v_0^2}}=
\sqrt{\bar{E}^2-1} 
\,;
\ee
then
\be
p\equiv \frac{1}{\bar{E}^2} = {1-v_0^2} \,.
\ee
The range of values of the parameter $p$ is $ 0 < p \leq 1 $, as for the 
Schwarzschild case, although now there is a 
difference: the 
observer starts out at $R=0$ instead of $R=\infty$ and it is outgoing 
instead 
of ingoing. In the de Sitter case one cannot start at $R=\infty$ 
because in the region $R> R_{H} \equiv H^{-1}$ beyond the  de Sitter 
horizon  
the 
geometry is not static and there is no conserved energy $E$ along 
timelike geodesics there. 

In principle, one could consider a radially ingoing timelike observer 
starting out at the de Sitter horizon $R_H $ with velocity $v_H \equiv 
-dR/d\tau \Big|_{R=H^{-1}} =-\bar{E}$, but the $\left( t, R\right) $ 
coordinates fail there. 

In terms of  the new time coordinate $\bar{T}$, the  line 
element~(\ref{dSstatic}) becomes
\be
ds^2=-fd\bar{T}^2 \pm 2 \sqrt{1-pf} \,d\bar{T} dR +p dR^2 +R^2 
d\Omega_{(2)}^2 \label{dSlinewithp}
\ee
with the upper [lower] sign referring to ingoing [outgoing] timelike 
geodesics. The metric is regular at the horizon $R_H=H^{-1}$: these 
coordinates penetrate the horizon and the time slices $d\bar{T}=0$ are not 
flat unless $p=1$.

The relation~(\ref{dSdT}) can be integrated explicitly to provide the new 
time 
\begin{eqnarray}
\bar{T} &= & T + \int dR \, \frac{\sqrt{1-pf}}{f} \label{useful-dS}\\
&&\nonumber\\
&=& T + \sqrt{1-p} \int 
dR \, \frac{ \sqrt{1+\frac{p}{1-p} \, H^2 R^2}}{ 1-H^2R^2} \,.
\end{eqnarray}
Using $\bar{H} \equiv \sqrt{ \frac{p}{1-p}} \, H$ and $\alpha \equiv 
\sqrt{\frac{1-p}{p} } \in \mathbb{R}$, one obtains
\begin{eqnarray}
\bar{T} &= &   T + \sqrt{1-p} \int dR \, \frac{  \sqrt{1+\bar{H}^2 R^2} 
}{
1-\alpha^2 \bar{H}^2R^2} \nonumber\\
&&\nonumber\\
&=& T + \frac{1}{\alpha^2 \bar{H}} \left[ 
\sqrt{\alpha^2+1} \tanh^{-1} \left( \frac{ \sqrt{\alpha^2+1}\, \bar{H}R}{ 
\sqrt{ 1+\bar{H}^2 R^2} }\right) -\sinh^{-1} \left( \bar{H}R \right) 
\right]\nonumber\\
&&\nonumber\\
&=& T + \frac{\sqrt{p} }{H} \left[ 
\frac{1}{p}  \tanh^{-1} \left( \frac{1}{ \sqrt{p(1-p)}} \, \frac{H R 
}{ 
\sqrt{1+  \frac{p}{1-p} \, H^2 R^2} } \right) -\sinh^{-1} \left( 
\sqrt{ \frac{p}{1-p}}\,  HR \right)  \right] \nonumber\\
&\,& + \mbox{const.}
\end{eqnarray}

\subsection{Painlev\'e-Gullstrand coordinates}

For the parameter value $p=1$, corresponding to vanishing initial velocity 
of the observer $v_0=0$, the line element~(\ref{dSlinewithp}) becomes
\be
ds^2=-fd\bar{T}^2 \pm 2HR \,d\bar{T} dR + dR^2 +R^2 
d\Omega_{(2)}^2  
\ee
(upper sign for ingoing, lower for outgoing geodesics), which is  the 
de Sitter line element in Painlev\'e-Gullstrand coordinates, which 
are therefore contained in the Martel-Poisson family of charts. Now the 
3-spaces of constant time $\bar{T}$ are Euclidean.

The Painlev\'e-Gullstrand time coordinate obtained from 
Eq.~(\ref{useful-dS}) for $p=1$ is
\begin{eqnarray}
\bar{T} &=& T+\int dR \, \frac{\sqrt{1-f}}{f} = T\pm \frac{1}{2H} \int 
dR \, \frac{HR}{1-H^2R^2} \nonumber\\
&&\nonumber\\
&=& T \pm \frac{1}{2H} \ln 
\left|1-H^2R^2\right| +\mbox{const.}
\end{eqnarray}
This is precisely the coordinate called ``Painlev\'e-de Sitter time'' used 
to study Hawking radiation with the tunneling method in 
Ref.~\cite{Parikh}.

\subsection{Eddington-Finkelstein coordinates}

EF coordinates for de Sitter space are used 
routinely, and they parallel the EF coordinates for Schwarzschild 
spacetime \cite{Eddington,Finkelstein}. The analogue of the tortoise 
coordinate is
\be
R^* \equiv \frac{1}{2H} \ln \left| \frac{1+HR}{1-HR}\right| = 
\frac{1}{2H} \ln \left| \frac{1-H^2R^2}{\left(1-HR\right)^2}\right| \,,
\ee
whose differential satisfies
\be
dR^* = \frac{dR}{1-H^2R^2} \,.
\ee
Null coordinates $\left(U,V\right)$ are introduced by 
\begin{eqnarray}
dU &=& dT-dR^* \,,\\
&&\nonumber\\
dV &=& dT+dR^* \,,
\end{eqnarray}
and 
\begin{eqnarray}
dT &=& dU+dR^*=dV-dR^* = \frac{dU+dV}{2} \,,\nonumber\\
&&\nonumber\\ 
dR^* &=& dV-dT= dT-dU = \frac{dV-dU}{2} \,. 
\end{eqnarray}
Although the parameter $p$ spans the 
range $\left( 0, 1 \right)$, one can formally obtain EF 
coordinates  by 
taking the limit $p\rightarrow 0$, which lies outside of this range, 
in the relevant equations. In 
this limit, the line element~(\ref{dSlinewithp}) becomes 
\be
ds^2 = -\left(1-H^2R^2\right) d\bar{T}^2 \pm 2d\bar{T}dR +R^2 
d\Omega_{(2)}^2 \label{dSinEF}
\ee
which is the well known de Sitter line element in EF 
coordinates ({\em e.g.}, \cite{Spradlin,mybook}), with the upper sign 
denoting EF coordinates based on ingoing null geodesics  and the lower 
sign denoting those based on outgoing null geodesics. 
In this limit the 
coordinate $\bar{T}$ (renamed $V$) is obtained by integrating in 
Eq.~(\ref{useful-dS}):
\begin{eqnarray}
V &\equiv & \lim_{p\rightarrow 0} \bar{T} = T + \int \frac{dR}{1-H^2R^2} 
\nonumber\\
&&\nonumber\\
&=& T + \frac{1}{H} \, \tanh^{-1} \left( HR \right) \nonumber\\
&&\nonumber\\
&=& T + \frac{1}{2H} \, \ln \left( \frac{1+HR}{1-HR} \right) \equiv T+R^*
\end{eqnarray} 
 and becomes the (null) advanced time. Introducing the retarded time 
as the second null coordinate
\be
U \equiv T - \frac{1}{2H} \, \ln \left( \frac{1+HR}{1-HR} \right) 
\equiv T-R^*\,,
\ee
the line element~(\ref{dSinEF}) can be written as
\begin{eqnarray}
ds^2&=& -\left(1-H^2R^2\right) dV\left(dV \mp 2dR^*\right) + R^2 
d\Omega_{(2)}^2 \\
&&\nonumber\\
&=& -\left(1-H^2R^2\right) \left[ dV^2 \pm dV(dV-dU )\right]
\end{eqnarray}
(upper sign for ingoing and lower for outgoing geodesics), where
\be
R\left(U,V \right) = \frac{1}{H} \, \tanh \left[ \frac{H\left(V-U 
\right)}{2} \right] \,.
\ee
Using
\begin{eqnarray}
dT&=&\frac{dU+dV}{2} \,,\\
&&\nonumber\\
dR&=& \frac{\left(1-H^2R^2\right)}{2} \left(dV-dU \right) \,,
\end{eqnarray}
one obtains
\be
ds_{(ingoing)}^2 = -\left(1-H^2R^2\right) dV\left(2dV-dU\right) +R^2 
d\Omega_{(2)}^2 
\ee
for ingoing null geodesics and
\be
ds_{(outgoing)}^2 = -\left(1-H^2R^2\right) dUdV +R^2 
d\Omega_{(2)}^2  
\ee
for outgoing null geodesics.
These line elements can be rewritten in terms of only one null coordinate 
$U$ or $V$, respectively, obtaining
\begin{eqnarray}
ds^2_{(+)} &=& -\left(1-H^2R^2 \right) dU^2 -2dUdR +R^2 d\Omega_{(2)}^2 
\,, \\
&&\nonumber\\
ds^2_{(-)} &=& -\left(1-H^2R^2 \right) dV^2 +2dVdR +R^2 d\Omega_{(2)}^2 
\,.
\end{eqnarray}

\section{Martel-Poisson charts for anti-de Sitter space}
\setcounter{equation}{0}
\label{sec:3}

Begin from the de anti-Sitter line element in curvature coordinates
\be
ds^2 =-\left(1+H^2R^2\right)dT^2 +\frac{dR^2}{1+H^2R^2} +R^2 
d\Omega_{(2)}^2 \equiv -fdT^2 +\frac{dR^2}{f} +R^2 
d\Omega_{(2)}^2  \label{AdSstatic}
\ee
and redefine the time coordinate $T\rightarrow \bar{T}$ according to 
\be
d\bar{T}= dT + \frac{ \sqrt{1-pf}}{f} \, dR \,. \label{AdSdT}
\ee
The equation of outgoing radial timelike geodesics  is again
\be
\frac{ds^2}{d\tau^2} = -f \left( \frac{dT}{d\tau} \right)^2 +\frac{1}{f} 
\, \left( \frac{dR}{d\tau} \right)^2 =-1   \label{dStimelikeradial}
\ee
and  a particle energy  is conserved along geodesics, $p_c T^c= -E $, 
giving 
\be
\left( \frac{dR}{d\tau} \right)^2= \bar{E}^2-f
\ee
and 
\be
\frac{dR}{d\tau} = \sqrt{ \bar{E}^2-f}
\ee
for outgoing geodesics. Introducing $ p\equiv 1/\bar{E}^2 $ 
and $v_0$ defined by 
\be
\frac{dR}{d\tau}\Big|_{R=0} = \sqrt{\bar{E}^2-1} =\gamma_0^2 v_0^2 \,;
\ee
the line element~(\ref{AdSstatic}) becomes 
\be
ds^2=-fd\bar{T}^2 \pm 2 \sqrt{1-p -pH^2R^2} \,d\bar{T} dR +p dR^2 +R^2 
d\Omega_{(2)}^2 \,.\label{AdSlineelement}
\ee
The Martel-Poisson coordinates are only defined for 
\be
0\leq R \leq  \sqrt{ \frac{1-p}{p}}  \, H^{-1} \equiv R_+ 
\,.\label{finiterange}
\ee 
In the limit $ p\rightarrow 1^{-}$ in which one expects to recover 
Painlev\'e-Gullstrand coordinates, $R_+\rightarrow 0$ and  this coordinate 
chart disappears.

If $0<p<1$, one can again obtain the time coordinate $\bar{T}$ in 
finite terms. 
Using the same $\bar{H} $ and $\alpha  $ as in the previous section, 
\begin{eqnarray}
\bar{T} &= & T+\int dR \, \frac{\sqrt{1-pf}}{f} \nonumber\\
&&\nonumber\\
&=& T+ \sqrt{1-p} \int dR \, \frac{ 
\sqrt{1-\frac{p}{1-p} \, H^2 R^2}}{
1+H^2R^2} \nonumber\\
&&\nonumber\\
&=& T+ \frac{\sqrt{1-p}}{\alpha^2 \bar{H}} \left[ 
\sqrt{\alpha^2+1} \tan^{-1} \left( \frac{ \sqrt{\alpha^2+1}\, \bar{H}R}{ 
\sqrt{ 1-\bar{H}^2 R^2} }\right) -\sin^{-1} \left( \bar{H}R \right) 
\right]\nonumber\\
&&\nonumber\\
&=& T + \frac{\sqrt{p} }{H} \left[ 
\frac{1}{p}  \tan^{-1} \left( \frac{1}{ \sqrt{p(1-p)}} \, \frac{H R 
}{ 
\sqrt{1- \frac{p}{1-p} \, H^2 R^2} } \right) -\sin^{-1} \left( 
\sqrt{ \frac{p}{1-p}}\,  HR \right)  \right] \nonumber\\
&\,& +\mbox{const.}
\end{eqnarray}

True PG coordinates for this metric, corresponding to 
the limit $p\rightarrow 1$, do not exist. In addition to the disappearance  
of the chart, $\Bar{T}$ becomes complex in this limit. This fact was noted 
in Ref.~\cite{LinSoo}. We come now to the crucial point, which is more 
general 
than the anti-de Sitter geometry. For completeness, before discussing 
this central issue, we report the EF coordinates for anti-de Sitter space.

\subsection{Eddington-Finkelstein coordinates}

One defines the tortoise coordinate $ r^*$ by imposing that the 
restriction of the metric to the 2-space $\left( T, r^* \right)$ is 
explicitly conformally flat, 
\be
-f dT^2 + f^{-1} dR^2 = f \left(-dT^2 + dr^{*2} \right) \,,
\label{confmetricAdS}
\ee
hence $ dr^* = dR/f = dR/\left( 1+H^2 R^2 \right)$ or, in finite form,
\be
r^* = \int{\frac{dR}{1 + H^2 R^2}} = \frac{\tan^{-1}{(HR)}}{H} \,.
\ee
The EF retarded and advanced times $ \left(u,v \right)$ 
are then  
\begin{eqnarray}
u & \equiv & T - r^* = T - \frac{\tan^{-1}{\left(HR \right)}}{H}\,,\\
& &\nonumber\\
v & \equiv &  T + r^* = T + \frac{\tan^{-1}{\left(HR \right)}}{H}
\end{eqnarray}
The outgoing and ingoing EF line elements follow by substituting $dT = du 
+ dr^*$ and $dT = dv - dr^*$ in the line element (\ref{AdSlineelement}), 
\begin{eqnarray}
ds_{(+)}^2  &=& -f du^2 - 2 \ du \, dR  + R^2 d\Omega_{(2)}^2 \,,
\label{uEFmetricAdS}\\
&&\nonumber\\
ds_{(-)}^2 &=& -f dv^2 + 2 \ dv \, dR + R^2 d\Omega_{(2)}^2\,.
\end{eqnarray}
Using  $dr^* = \left(dv - du \right)/2$  in Eq.~(\ref{uEFmetricAdS}) 
yields
\be
ds^2 = - f \ dudv + R^2 d\Omega_{(2)}^2 \,.
\ee

\section{PG coordinates and Misner-Sharp-Hernandez mass}
\label{sec:4}
\setcounter{equation}{0}

A rather general recipe to construct PG coordinates for any spherically 
symmetric metric (static or not) is given in Ref.~\cite{NielsenVisser}. 
Begin with the line element in the Abreu-Nielsen-Visser gauge 
\cite{NielsenVisser, AbreuVisser} 
\be
ds^2 =- \mbox{e}^{2\Phi(t,R)} \left( 1-\frac{2M_\text{MSH}(t,R)}{R} 
\right) dt^2 
+\frac{dR^2}{ 1-2M_\text{MSH}(t,R)/ R } +R^2 d\Omega_{(2)}^2 
\label{ANVgauge}
\ee
employing the areal radius $R$ as the radial coordinate. Here 
$M_\text{MSH}(t,R)$ 
is the Misner-Sharp-Hernandez mass well known in spherical fluid 
mechanics and in gravitational collapse \cite{MSH1,MSH2}. (It is not 
trivial that this is 
the object appearing in Eq.~(\ref{ANVgauge})---see \cite{NielsenVisser, 
AbreuVisser} for an explanation.)

Define the new time coordinate $\bar{t}\left( t, R\right)$ by
\be
d\bar{t} = \frac{\partial \bar{t}}{\partial t} \, dt 
+ \frac{\partial \bar{t}}{\partial R} \, dR \equiv \dot{\bar{t}} dt 
+\bar{t}' dR \,.
\ee
substituting into the line element and requiring $g_{RR}=1$ leads to 
\cite{NielsenVisser}
\be
\bar{t}'=\pm \frac{ \sqrt{ 2M_\text{MSH}/R}}{1-2M/R} \, \mbox{e}^{\Phi} \, 
\dot{\bar{t}} \,,
\ee
which has always a solution. Then the line element in PG coordinates taje 
the form
\be
ds^2 = - \left[ c^2\left(\bar{t}, R \right) -v^2\left( \bar{t}, R \right) 
\right] d\bar{t}^2 + 2 v\left( \bar{t}, R \right)  d\bar{t}dR +R^2 
d\Omega_{(2)}^2 \,,
\ee
where 
\begin{eqnarray}
c\left( \bar{t}, R \right) &=& \frac{ \mbox{e}^{-\Phi} }{\dot{\bar{t}}} 
\,,\\
&&\nonumber\\
v \left(\bar{t}, R \right) &=& c\left(\bar{t}, R \right) \, \sqrt{ 
\frac{2M_\text{MSH}}{R} } \,.
\end{eqnarray}
In practice, the function $\bar{t}\left( t, R \right) $ is not always 
determined explicitly. This is equivalent to introducing an integrating 
factor to make $d\bar{t}$ an exact differential \cite{mybook}.

It is clear that the Nielsen-Visser procedure breaks down in regions where 
the mass $M_\text{MSH}$ becomes negative and $v$ becomes imaginary.  This 
is exactly 
the case of anti-de Sitter space in the region $0\leq R<H^{-1}$ covered by 
the locally static coordinates, and of the inner region of the 
Reissner-Nordstr\"om spacetime pointed out in~\cite{LinSoo} (although the 
procedure of \cite{NielsenVisser} to construct PG coordinates is not 
mentioned there). Trivial as it may seem, this observation explains from 
the {\em mathematical} point of view why one cannot construct PG 
coordinates in these two spaces and, more in general, in any region with 
negative Misner-Sharp-Hernandez mass.

Let us come now to the {\em physical} explanation. As usual, the 
Schwarzschild spacetime taken as an example sheds light on other 
geometries. Consider the Schwarzschild spacetime~(\ref{Schwarzschild})  
with negative mass, which has a naked central singularity and no horizons. 
The Misner-Sharp-Hernandez is $M_\text{MSH}=-|m|<0$ and PG coordinates 
cannot 
be constructed. The reason is that these coordinates are associated with 
observers falling in radially from infinity with zero initial velocity. 
Since gravity is now repulsive, these particular observers cannot even 
begin to fall because they cannot overcome the repulsion and must move 
outwards instead. There are no ingoing timelike radial geodesics with zero 
initial velocity. To wit, repeat the procedure of Sec.~\ref{sec:1} to 
obtain, along radial timelike geodesics,
\be
\left( \frac{dr}{d\tau} \right)^2 =\bar{E}^2-f=\bar{E}^2 -1 
-\frac{2|M|}{r} \,;
\ee
imposing zero initial velocity at infinity gives
\be
v_{\infty}^2 = \left( \frac{dr}{d\tau} \right)^2 \Big|_{\infty}= 
\bar{E}^2-1=0
\ee
or $\bar{E}=1$. Then at any radius $r\in \left(0, +\infty \right)$ it is
\be
\left( \frac{dr}{d\tau} \right)^2 = -\frac{2|M|}{r} <0 \,,
\ee
whic is clearly impossible. Therefore, PG observers cannot be defined 
because of the repulsion. The situation is the same in anti-de Sitter 
space, except that now the observer starts at the centre. We have again 
(changing $f \rightarrow 1+H^2R^2$),
\be
\left( \frac{dr}{d\tau} \right)^2 = \bar{E}^2-f = \bar{E}^2-1-H^2R^2 
\ee
and, imposing that the initial velocity at the centre vanishes,
\be
v_0^2 \equiv \left( \frac{dR}{d\tau} \right)^2 \Big|_{R=0} =\bar{E}^2-1=0 
\,,
\ee
one obtains again $\bar{E}=1$ and
\be
\left( \frac{dr}{d\tau} \right)^2 =-H^2R^2 <0  
\ee
for all $R\in \left( 0, H^{-1} \right)$, which clearly shows the 
impossibility of defining PG observers. This is due to the fact that the 
negative cosmological constant repels and confines a particle at the 
centre. If the particle has zero initial velocity there, it will not exit. 
By contrast, the positive cosmological constant of de Sitter space 
attracts a particle located at $R=0$ toward larger and larger values of 
$R$.

\section{Einstein static universe}
\setcounter{equation}{0}
\label{sec:5}

In general relativity, the static Einstein universe \cite{Einstein} arises 
from the delicate balance between a dust and the positive cosmological 
constant, and is unstable with respect to homogenous perturbations 
\cite{Eddington2}. Stability with respect to vector and tensor 
perturbations is a different issue, and stability with respect to 
inhomogeneous scalar density perturbations depends on the sound speed 
$c_s$ \cite{Harrison,Gibbons,BarrowEllisMaartensTsagas}, with neutral 
stability occurring if $c_s>1/sqrt{5}$, a range that also maximizes 
entropy \cite{Gibbons}.
 
Modern interest in this solution arises in braneworld models 
\cite{8,9,10,11}, loop quantum cosmology \cite{13,14}, string theory 
\cite{12}, analog gravity \cite{Volovik}, with generalizations to 
non-constant pressure 
\cite{15,16,17,18,19,20}. Further motivation for the study of the static 
Einstein universe comes from the possibility that the early inflationary 
universe might have begun in an asymptotic Einstein state 
\cite{EllisMaartens}. Moreover, the static Einstein universe has seen 
renewed attention as a solution of the field equations of modified gravity 
theory \cite{ESU}. Our considerations in this section are independent of 
the theory of gravity.

For the positively curved Einstein static universe, introducing the 
Martel-Poisson coordinates proceeds as outlined in \cite{MartelPoisson}. 
The line element is 
\be
ds^2=-dt^2+a_0^2 \left( \frac{dr^2}{1-r^2} +r^2 d\Omega_{(2)}^2 \right) 
\,,\label{SEU}
\ee
where $0\leq r<1$. This geometry has the timelike Killing vector $t^a = 
\left( \partial 
/\partial t 
\right)^a $ and areal radius $R=a_0r$. The energy $E$ of a test particle  
is conserved along the geodesic. Along radial timelike geodesics, 
$dt/d\tau=\bar{E} \equiv E/m$ and, substituting into the normalization 
$u_c u^c=-1$ yields 
\be
\left( \frac{dr}{d\tau}\right)^2 = \frac{\bar{E}^2-1}{a_0^2}\, \left( 
1-r^2 \right) \,.
\ee
The proper 3-velocity at $r=0$ has magnitude
\be
v_0=\left| \frac{dr}{d\tau} \right| = \frac{ \sqrt{\bar{E}^2-1}}{a_0} 
\ee
so that the parameter $p$ is again
\be
 p \equiv \frac{1}{\bar{E}^2} = 1-v_0^2\,, 
\ee
it has the same meaning as in the de Sitter universe, and it spans the 
range $0< p \leq 1$. Defining the new time coordinate 
$\bar{t}$ by \cite{MartelPoisson}
\be
d\bar{t}=dt + \sqrt{ \frac{1-p}{1-r^2} } \, dR \,,\label{dbart}
\ee
the line element becomes
\be
ds^2 = -d\bar{t}^2 +2\sqrt{ \frac{1-p}{1-r^2} } \, d\bar{t} dR 
+\frac{ pdR^2}{1-r^2}  +R^2 d\Omega_{(2)}^2 \,.\label{SEUp}
\ee
Using $\alpha \equiv \sqrt{ (1-p)/p}$, the integration of 
Eq.~(\ref{dbart}) gives
\begin{eqnarray}
\bar{t} &=& t + a_0 \sqrt{1-p} \int \frac{dr}{ \sqrt{1 - r^2} }  
\nonumber\\
&&\nonumber\\
&=& t+ a_0 \sqrt{1-p} \arcsin r + \mbox{const.}
\equiv t+ a_0 \sqrt{1-p} \, \chi + \mbox{const.}\,,
\end{eqnarray}
where $\chi$ is the usual hyperspherical radius \cite{Wald}. The proper 
radius $a_0\chi$ (which could also be called ``volume'' radius) is 
distinct from the areal radius $R$ in spatially curved FLRW universes. 

\subsection{PG coordinates}

By taking the limit $p\rightarrow 1 $, $d\bar{t}$ reduces to $dt$ in 
Eq.~(\ref{dbart}), and the line element~(\ref{SEUp}) 
reverts to the static FLRW line element~(\ref{SEU}) in comoving 
coordinates, in which the spatial sections are positively curved. 
Again, setting $v_0=0$ implies $\bar{E}=1$ and $\left( dr/d\tau \right)^2 
<0$ along radial timelike geodesics. PG coordinates cannot be introduced 
as a limit of the Poisson-Martel family of charts.  The reason is rather 
simple: since the matter content of this universe is 
dust and its collapse is (just) balanced by the positive cosmological 
constant, a test particle with zero radial initial velocity, {\em i.e.}, 
initialy comoving with the cosmic substratum, remains comoving 
with it---that is, not moving 
at all. Massive particles on timelike radial geodesics need nonzero 
initial velocity to move.

PG coordinates can still be introduced following the procedure of 
\cite{NielsenVisser,AbreuVisser}, which yields $ 
M_\text{MSH}(R)=R^3/\left(2a_0^2\right) $ and 
\be
c\left( \bar{t},R \right) = \frac{a_0}{R} \, \bar{t}' \,,
\ee
\be
v\left( \bar{t},R \right) = \pm \frac{2a_0}{R} \, \bar{t}'   \,,
\ee
and the line element in PG coordinates is 
\be
ds^2=-\frac{a_0^2}{R^2} (\bar{t}')^2 \left(1-\frac{R}{a_0} \right) 
d\bar{t}^2 \pm \frac{2a_0 \bar{t}'}{R}\, d\bar{t}dR + R^2 d\Omega_{(2)}^2 
\,.
\ee 
The Martel-Poisson interpretation of PG coordinates does not apply to the 
static Einstein universe.

\subsection{Eddington-Finkelstein coordinates}

The tortoise coordinate $r^*$ is defined so that 
$ dr^* = dR/\sqrt{ f} $ and, integrating,
\be
r^* =  \int{\frac{dR}{\sqrt{1-\left(\frac{R}{a_0} \right)^2}}} = a_0 
\sin^{-1}{r} = a_0 \chi \,,
\ee
where $\chi $ is the usual hyperspherical radius \cite{Wald}. The retarded 
and advanced times are now
\begin{align}
u & \equiv t - r^* = t - a_0 \sin^{-1}{r} \,,\\
&\nonumber\\
v & \equiv t + r^* = t + a_0 \sin^{-1}{r} \,.
\end{align}
 As expected, $v=\lim_{p\rightarrow 0} \bar{t}$.

With the substitutions $dt = du + dr^*$ and $dt = dv - dr^*$,  the 
outgoing/ingoing EF line 
elements are 
\begin{align}
ds_{(+)}^2 &  =  -du^2 - \frac{2 dudR}{\sqrt{1 - r^2}}+ R^2 
d\Omega_{(2)}^2\,,\\
&\nonumber\\
ds_{(-)}^2 & =  -dv^2 + \frac{2 dvdR}{\sqrt{1 - r^2}}  + R^2 
d\Omega_{(2)}^2 \,.
\end{align} 
Then, using  $dr^* = \left(dv - du\right)/2$, one 
obtains
\begin{align}
ds^2 & =  -dudv + R^2 d\Omega_{(2)}^2 \,. \label{uvSEmetric}
\end{align}

\section{Conclusions}
\label{sec:6}
\setcounter{equation}{0}

The PG coordinates originally introduced for 
the Schwarzschild geometry \cite{Painleve,Gullstrand} have proved very 
useful in the study of study the thermodynamics of 
black holes and of other horizons, especially in the context of the 
tunneling formalism of Parikh and Wilczek \cite{ParikhWilczek,Parikh}. It 
is rather 
unfortunate that this coordinate chart cannot be introduced for the most 
important space of string theories, anti-de Sitter space  associated with 
a negative cosmological constant, and for the Schwarzschild-anti de Sitter 
geometry obtained by embedding the Schwarzschild black hole into it. This 
difficulty has been noted, but not explained, in the literature and its 
physical intepretation has remained a puzzle. We have clarified this 
anomaly by 
looking at the physical meaning of PG observers in static cosmological 
spacetimes. While, in asymptotically flat spherical spacetimes, PG 
observers fall in radially from infinity, starting with zero initial 
velocity, in cosmological settings instead they fall outward fromm $R=0$. 
In anti-de Sitter space, where the Misner-Sharp-Hernandez quasilocal 
mass is negative and repulsive because of the negative cosmological 
constant, a would-be PG observer starting at the 
centre  with zero initial velocity cannot overcome this repulsion and move 
away. Similarly, in the Schwarzschild spacetime with negative mass, an 
observer located at infinity with zero initial velocity does not fall 
radially toward smaller radii because it is repelled by the negative mass 
at the central singularity. 

This physical interpretation applies to generic  regions containing 
negative Misner-Sharp-Hernandez mass, which repels instead of attracting. 
Martel-Poisson observers different from PG ones, and Lorentz-boosted with 
respect to them, start out radially with non-vanishing initial velocity 
and have a chance to overcome the initial repulsion, at least for part of 
their journey before they are turned around by repulsion, which causes 
Martel-Poisson coordinates to have a range smaller than the entire locally 
static region (cf. Eq.~(\ref{finiterange}) for anti-de Sitter space).

In the case of (non-extremal) Schwarzschild-(anti)-de Sitter black holes, 
where there are two horizons, radial timelike geodesics cannot start ar 
$R=0$ not at $R=\infty$. In this case it is more convenient to start 
somewhere in between, but then the physical meaning of the Martel-Poisson 
coordinates is altered. This situation will be discussed in a separate 
work.

\small \section*{Acknowledgments} This work is supported by the 
Natural Sciences \& Engineering Research Council of Canada (Grant no. 
2016-03803) and by Bishop's University.

\end{document}